\documentclass[11pt]{article}
\usepackage{graphicx}
\begin{document}
~~~\\
\begin{center}
{\bf  Decoherence and dissipation in quantum two-state systems \\[1cm]}
{ M. Grigorescu \\[1cm]}
\end{center}

\noindent
$\underline{~~~~~~~~~~~~~~~~~~~~~~~~~~~~~~~~~~~~~~~~~~~~~~~~~~~~~~~~
~~~~~~~~~~~~~~~~~~~~~~~~~~~~~~~~~~~~~~~~~~}$ \\[.1cm]
{\bf Abstract} \\
{\small The Brownian dynamics of the density operator for a quantum system 
interacting with a classical heat bath is described using a stochastic, 
non-linear Liouville  equation obtained from a variational principle. 
The rate of the noise-induced transitions is expressed as a
function of the environmental spectral density,  and  is discussed for the 
case of white noise and blackbody radiation. The time-scales of
decoherence and dissipation are investigated numerically for a system
of two quantum states. These  are the ground and first excited state of the 
center of mass  vibrations for an ion confined in a harmonic trap.} 
$\underline{~~~~~~~~~~~~~~~~~~~~~~~~~~~~~~~~~~~~~~~~~~~~~~~~~~~~~~~~
~~~~~~~~~~~~~~~~~~~~~~~~~~~~~~~~~~~~~~~~~~}$ \\[.5cm]
{\bf I. Introduction}  \\[.5cm] \indent 
The evolution of a quantum system surrounded by a thermal environment
represents a fundamental problem of the quantum theory,
increasingly important for applications.
In classical mechanics, a particle interacting with a heat bath
has a Brownian evolution towards thermalization, which reduces
to a purely dissipative motion when the temperature decreases to
zero \cite{hang}.  For a quantum particle, the interaction with the 
environment is more complex and may produce phenomena without a classical 
analog, as the collapse  of the wave function  and  decoherence \cite{zurek}. 
These effects are particularly
important for understanding the mechanism of atom switching in the scanning 
tunneling microscope \cite{stm}, for  the practical realization of a quantum 
computer \cite{qc} \cite{pell}, or the transmission of quantum information  
\cite{bennett}. However, despite the growing interest in these 
fields, the physics of the Brownian quantum dynamics and  of the
wave function decoherence is not yet completely understood \cite{sipe}
\cite{koba}.  \\ \indent
The conceptual difficulties appearing in the description of a quantum system 
interacting with a classical environment are well illustrated 
by  the "Schr\"odinger cat experiment", or the Zeno paradox at the
continuous measurement process \cite{2} \cite{3}. The 
wave function collapse at macroscopic scale (e.g. of the apparatus 
pointer) was investigated by  Zurek \cite{zurek}, assuming  that  
quantum mechanics describes both the environment and the system.
However, when only the system is quantum, the reduction of the density
operator by partial tracing over the environment's degrees of freedom
is not possible, and the approach should be different. 
The problem occurs, for instance, in quantum gravity, whenever the 
space-time framework of a quantum particle remains classical \cite{rrp}.
In such cases, a promising approach to the mixed classical-quantum
dynamics consists of using 
coupled  Hamilton-Heisenberg equations derived from a variational principle 
\cite{aa}.  The time-dependent variational principle (TDVP) \cite{kramer} is
frequently used in the quantum many-body theory to obtain
the mean-field equations  (e.g. time-dependent Hartree-Fock).
The mean-field dynamics appears by kinematical constraints on the unitary 
evolution in the many-body Hilbert space, and has a 
mixed character, including both quantum and classical aspects \cite{rjp93}. 
Therefore, the TDVP provides an appropriate frame to treat also the mixed
version of the bilinear coupling  model, consisting of a quantum particle 
interacting with a heat bath of classical harmonic oscillators.
For the pure states, this approach leads  to a non-linear time-dependent 
Schr\"odinger equation, containing additional noise and friction terms 
\cite{gc}.  However, for the treatment of the thermalization process, and 
evolution from pure to mixed states, it is necessary to include the
interaction with the environment in the dynamical equation for the density 
matrix \cite{pb96}.  \\ \indent
In this work, the combined effect produced by noise and dissipation
on the quantum evolution will be described using a stochastic, 
non-linear Liouville equation. This equation  can be derived from the 
variational principle, and will be presented in Sec. II.  \\ \indent
The noise represents an external stochastic force which stimulates
transitions between the stationary states  of the quantum system. A 
perturbative calculation of the transition rates is presented in Sec. III.  
The result is compared with the one obtained when the environment is quantized,
and is illustrated by two examples: the  Gaussian 
white-noise and the thermal  radiation.  \\ \indent
The evolution of the density operator when noise and
dissipation are both included is investigated numerically in Sec. IV,
for the case of a quantum two-state system with relevance for the operation of
a logical  gate in a  quantum computer. This system consists
of the ground and first excited state of the center of mass vibrations for a 
cooled ion  in a harmonic trap. \\ \indent
The summary of the results and the conclusions are  presented in Sec. V.
\\[.5cm]
{\bf II. The stochastic Liouville equation } \\[.5cm] \indent
Let us  consider a quantum system with the Hamiltonian and density 
operators $H_0$ and $\rho$, respectively.   These operators are 
Hermitian, and their matrices in an orthonormal basis $\{ \vert e_n \rangle \}$ can be 
diagonalized by unitary transformations. If $\vert \Psi_n\rangle (t)$ denotes
a normalized eigenstate of $\rho (t)$ corresponding to the eigenvalue
$w_n$, then there exists an unitary operator $\chi$, such that\footnote{In the
finite-dimensional case we may consider also the Gauss decomposition  $\rho =  
\chi  D  \chi^\dagger $ where the matrices $\chi$ form a nilpotent group, 
and $D$ is diagonal.} $\rho =  \chi  D  \chi^\dagger $, with 
 $ D= \sum_n w_n \vert e_n\rangle \langle  e_n \vert$. The operator $ D$
 is related to the von Neumann entropy \cite{vn} $S = - Tr( \rho \ln
\rho) = - Tr( D \ln D)=  - \sum_n w_n \ln w_n$. 
According to Reznik \cite{reznik} 
it is convenient to introduce a "square root operator" $\eta$,
such that $\rho= \eta \eta^\dagger$. This
operator can be defined as $\eta = \chi \sqrt{
{ D}}$,  and in this case $D = \eta^\dagger \eta= \rho - 
[\eta, \eta^\dagger]$. 
\\ \indent
Following the assumptions of the bilinear coupling model \cite{zw} \cite{cl},
a thermal environment can be simulated by a classical heat 
bath of harmonic oscillators. These will be supposed to be continuously 
distributed in frequency with the density $\mu_\omega$, and interacting with 
the quantum system
by the Hamiltonian ${ H}_{coup} =  { K} \int_0^\infty d \omega \mu_\omega 
C_\omega q_\omega$. Here ${ K}$  denotes a Hermitian 
coupling operator,  $C_\omega$ are constants, 
and $q_\omega$ are the  time-dependent bath coordinates.  
The evolution of this mixed  classical-quantum system 
will be described using the variational equation 
\begin{eqnarray}
\delta \int dt \{ \int_0^\infty d \omega \mu_\omega (  
\dot q_\omega p_\omega  -  h_\omega )  
+  Tr \lbrack \eta^\dagger i \hbar \partial_t \eta - 
\eta^\dagger ({ H_0 }+ { H}_{coup}) \eta  \rbrack \} =0~~,
\end{eqnarray}
where the dynamical variables of the quantum system are $\eta$, $\eta^\dagger$, 
and $h_\omega = (p_\omega^2  + m_\omega^2  \omega^2
q_\omega^2 )/2 m_\omega$ is the classical Hamiltonian for a bath oscillator.   
The equations of motion obtained from  Eq. (1) at the variation of   
 $\eta (t)$, $\eta^\dagger (t)$,
 $q_\omega (t)$ and $p_\omega (t)$ are 
\begin{equation}
i \hbar \partial_t \eta = ({ H_0} 
 + { K} \int_0^\infty d \omega \mu_\omega C_\omega q_\omega )
\eta~,~ 
i \hbar \partial_t \eta^\dagger = - \eta^\dagger (H_0 
 + { K}  \int_0^\infty d \omega \mu_\omega C_\omega q_\omega )
\end{equation}
for the quantum system, and
\begin{equation} \dot q_\omega = \frac{p_\omega}{m_\omega}~~,~~
\dot p_\omega = - m_\omega \omega^2 q_\omega - C_\omega Tr( { K} \rho)
\end{equation}
for the classical oscillators. The system of Eq. (2) leads to the
 Liouville equation for the density operator,
\begin{equation} 
i \hbar \partial_t \rho = [{ H_0} 
 + { K} \int_0^\infty d \omega \mu_\omega C_\omega q_\omega ,  \rho] ~~. 
\end{equation}
The classical equations (3) can be solved in terms of the unknown function 
of time $Q_\rho (t) \equiv Tr(  K \rho (t) ) $, and 
their retarded solution is 
$$
q_\omega (t) = [q_\omega (0) + g_\omega Q_\rho (0) ]
\cos \omega t + \frac{ {\dot q}_\omega (0) }{\omega} \sin \omega t 
$$
\begin{equation}
- g_\omega  Q_\rho (t) + g_\omega \int_0^t dt'   
\dot{ Q}_\rho (t')  \cos \omega (t-t')  ~~,
\end{equation}
with $g_\omega = {C_\omega}/({m_\omega \omega^2 })$.
When  this solution is inserted in Eq. (4),  the  Liouville equation
becomes
\begin{equation}
i \hbar \partial_t  \rho = [{ H}_0 + { W}(t) , \rho]
\end{equation} 
where ${ W}(t) = - { K} [ \xi(t) + f_\rho (t)+ Q_\rho (t) \Gamma(0)-Q_\rho (0) \Gamma(t)] $.  
The additional term $W$ contains the force functions $\xi(t)$ and $f_\rho (t)$,
\begin{equation}
\xi (t)=  
- \int_0^\infty d \omega \mu_\omega C_\omega \lbrack q_\omega (0) \cos \omega t  
+ \frac{  {\dot q}_\omega (0)}{\omega} 
\sin \omega t \rbrack~~,
\end{equation}
\begin{equation}
f_\rho (t) = - \int_0^t \Gamma(t-t' ) \dot{Q}_\rho (t') dt'~~,~~
\Gamma (t) = \frac{2}{\pi} \int_0^\infty d \omega J(\omega)  \frac{
\cos \omega t}{\omega} ~~,
\end{equation}
where $J(\omega) = \pi \omega g_\omega C_\omega \mu_\omega/2$
is the spectral density of the environment.  
The operator  $K[Q_\rho (t) \Gamma(0) -Q_\rho (0) \Gamma(t)]$ ensures the invariance
of $W$ to the renormalization of $\Gamma$ by an additive constant, and vanishes when
$\Gamma (t) \sim \delta (t)$ or $Q_\rho(t)=0$. In the case of a classical heat bath at
thermal equilibrium we will assume that $Q_\rho$ and $\Gamma$ are such that
$Q_\rho (t) \Gamma(0) -Q_\rho (0) \Gamma(t)=0$. \\ \indent
The force function $f_\rho$ makes Eq. (6) non-linear, and therefore, 
in general, the solutions $\rho (t)$ will not satisfy the superposition
principle. Though, exceptions may occur on distinguished
"superselection sectors", as the trivial one defined by the set 
$\{ \rho / \dot{Q}_\rho (t) \equiv 0, t \in [0, \infty) \}$. 
\\ \indent
If the initial coordinates and momenta $q_\omega (0)$, $ p_\omega (0)$ 
are distributed within a statistical ensemble
${\cal E}$ with the temperature $T$, then  
$\xi(t)$ behaves like a stochastic force which has zero mean and is
related to the memory function $\Gamma(t)$ by 
the fluctuation-dissipation theorem (FDT). Denoting by $ \ll  * \gg$
the average over ${\cal E}$, these properties of $\xi$ are expressed by
$ \ll  \xi(t) \gg =0$ and  $  \ll  \xi(t) \xi(t') \gg =  k_B T \Gamma(t-t') $,
respectively. In this case, Eq. (6) becomes a  stochastic
non-linear Liouville equation for the density operator,
\begin{equation}
i \hbar \partial_t \rho = [{ H}_0  
- { K} (\xi(t) + f_\rho (t) ) , \rho]~~.
\end{equation} 
 Along the Brownian trajectories $\rho{(t)}$ obtained by solving this 
equation,  $Tr(\rho^k)$ is a constant for every
power $k=1,2...$, and therefore the density operator 
has a Brownian evolution  which preserves the "purity" of the initial state.
However, decoherence may appear for the average density operator 
$\rho_{av} (t)  \equiv  \ll  \rho\gg (t)= \sum_{r=1,N_t} \rho^r {(t)}/N_t$  
calculated over an ensemble of $N_t$ trajectories $\rho^r {(t)}$ generated 
with the same initial condition, $\rho^r \vert_{t=0} = \rho_0$. 
This average (supposed to be equivalent to the average over ${\cal E}$)
accounts for the dynamics of the occupation probabilities,
and should satisfy a quantum transport equation    
 \cite{zurek,cl,hove,dekker}.  
\\[.5cm]
{\bf III. The noise-induced transitions } 
\\[.5cm] \indent
When the non-linear friction force $f_\rho$ vanishes or it can be neglected,  
Eq. (9) becomes
\begin{equation}
i \hbar \partial_t \rho = [{ H}_0 + H_{noise}(t) , \rho]~~,
\end{equation} 
where $H_{noise} (t)= -{ K} \xi(t)$ is the noise term  due to the 
fluctuating environmental forces. This equation  has the solution
\begin{equation}
\rho(t) = e^{- i  H_0 t / \hbar} \tilde{ \rho} (t) e^{i H_0 t /
\hbar}~~,~~
\tilde{\rho}_{(t)} = {\cal T} e^{ \frac{i}{\hbar} \int_0^t dt' 
\xi (t') {\cal L}_{\tilde{K} (t')} }~ \rho_{0} ~~
\end{equation}
where ${\cal T}$ denotes the time-ordering operator \cite{bd}, 
\begin{equation}
\tilde{K} (t) =  e^{ i  H_0 t / \hbar} K e^{-i H_0 t /
\hbar}
\end{equation}
 is the coupling operator in the interaction representation,  and  
${\cal L}_A$ is the Lie derivative with respect to the operator $A$
defined by the commutator, ${\cal L}_A B  \equiv [A,B]$. 
The ensemble average of $\rho (t)$ can be written as
\begin{equation}
\rho_{av} (t)= e^{- i  H_0 t / \hbar} \tilde{ \rho}_{av} (t)
e^{i H_0 t / \hbar}~~,
\end{equation}
and the terms appearing in the expansion of the
time-ordered exponential defining $\tilde{ \rho}_{av} (t)$
can be calculated using the FDT. Retaining only the 
first non-vanishing average, the result is
\begin{eqnarray} 
\tilde{ \rho}_{av}(t) = \rho_0  -  \frac{ k_B T}{\hbar^2}
\int_0^t dt_1 \int_0^{t_1} dt_2 \Gamma(t_1 
-t_2) [ \tilde{K} (t_1), [ \tilde{K} (t_2), \rho_0 ] ] ~~.
\end{eqnarray}
This formula can be used to estimate the rate of the noise-induced transitions
 between the energy eigenstates $\{ \vert E_k\rangle \}$ defined by $H_0 \vert E_k\rangle = 
E_k \vert E_k\rangle$. If initially the system is in the pure state
$\vert E_i\rangle$, then $\rho_{0} = \vert E_i\rangle \langle E_i \vert$, and the rate of the  
transition $\vert E_i\rangle \rightarrow  \vert E_f\rangle$ can be defined by the asymptotic
time-derivative 
\begin{equation}
\lambda_{fi} =
 \frac{d v_f}{dt} \vert_{t \rightarrow \infty}
\end{equation}
of the occupation probability
$
v_f (t)=  \langle  E_f \vert \rho_{av} (t) \vert E_f\rangle=   \langle  E_f \vert \tilde{ \rho}_{av} (t) 
\vert E_f\rangle $. Using Eq. (14), this probability is 
\begin{eqnarray}
v_f (t) = 2  k_B T
\frac{ \vert { K}_{fi} \vert^2 }{ \hbar^2} 
\int_0^t dt_1 \int_0^{t_1} dt_2 \Gamma(t_1 
- t_2) \cos[ \Omega_{fi} (t_1-t_2)]
\end{eqnarray}
where ${ K}_{fi} =  \langle  E_f \vert { K}  \vert E_i \rangle $ is the matrix
element of the coupling operator and $\Omega_{fi}= (E_f - E_i)/ \hbar$ . 
The  memory function appearing here can be expressed in terms of $J(\omega)$ 
using Eq. (8), and  the transition rate takes the form
\begin{equation}
\lambda_{fi}^{ noise}  =  \frac{2  }{\hbar^2} \vert { K}_{fi} 
\vert^2 k_B T
\frac{ J( \omega_{fi})}{\omega_{fi}}~~,~~ \omega_{fi} = \vert \Omega_{fi}
\vert.
\end{equation}
\indent
This result was obtained assuming a classical environment,
but such cases are rare. Therefore, it is interesting
to compare Eq. (17) with the average transition  rate $\lambda_{fi}^Q$ 
determined by the whole coupling interaction ${ H}_{coup}$ when
the bath oscillators are quantized. Let us denote by $\vert n\rangle$
the eigenstates of Hamiltonian operator
$\hat{h}_\omega$ and by ${\hat \rho}_\omega$ the related density operator.
At thermal equilibrium, ${\hat \rho}_\omega = Z_\omega^{-1}
\exp( - \hat{h}_\omega/k_B T) $ with
$Z_\omega = Tr[ \exp( - \hat{h}_\omega/k_B T)]$. The correspondent
of the trajectory average $\rho_{av}$ is
$\rho^{Q}_{av} (t) \equiv Tr_{\cal E}[ {\cal R} (t)]$,  defined by the partial 
trace over the environmental Hilbert space of the density operator ${\cal R}$ 
for the whole system, and within the same approximation as above, 
\begin{eqnarray}
v_f(t) = - \frac{1 }{\hbar^2}
\int_0^t dt_1 \int_0^{t_1} dt_2  \langle E_f \vert  
Tr_{\cal E} \{ [  \tilde{H}_{coup} (t_1),  [ \tilde{H}_{coup} (t_2), 
{\cal R}_0 ] ] \} \vert E_f\rangle ~.
\end{eqnarray}
The interaction representation for $H_{coup}$ is defined here with respect
to the total Hamiltonian ${\cal H}_0 =H_0 + H_{\cal E}$,
$H_{\cal E} = \int_0^\infty d \omega \mu_\omega
\hat{h}_\omega$, and 
\begin{equation}
{\cal R}_0 = \vert E_i\rangle \langle E_i \vert \otimes Z^{-1}_{\cal E} \exp(- H_{\cal E}/k_B T)~~,
\end{equation}
with $Z_{\cal E}$ the partition function of the environment. 
For an excitation process $\vert E_i\rangle  \rightarrow \vert E_f\rangle$,  Eq. (15) 
with $v_f$ of Eq. (18) gives the rate   
\begin{eqnarray}
\lambda^{ \uparrow Q} = \frac{ 2 \pi}{\hbar^2} \vert { K}_{fi} \vert^2
\int_0^\infty d \omega \mu_\omega \sum_{n=1}^\infty 
({\hat \rho}_\omega)_{nn}  
C_\omega^2
 \vert (\hat{q}_\omega)_{n-1,n}  \vert^2 \delta( \omega_{fi} - \omega)~~,
\end{eqnarray}
consisting of the ensemble average of the Fermi's golden rule \cite{bend}.
This average can be written in the compact form 
\begin{equation}
\lambda^{ \uparrow Q}=   A_{fi} \langle n\rangle_{\omega_{fi}}~~,~~ 
A_{fi} = \frac{2  }{\hbar} \vert { K}_{fi} \vert^2 
 J(\omega_{fi})  ~~,
\end{equation}
where $A_{fi}$ denotes the rate of the (non-thermal) spontaneous decay, and
$ \langle n\rangle_\omega= 1 /( e^{ \hbar \omega / k_B T} -1)$ is the average
number of the bath phonons with energy $\hbar \omega$.
Similarly, the total rate for a decay process $\vert E_f\rangle \rightarrow  \vert E_i\rangle$ 
is $\lambda^{ \downarrow Q}=   A_{fi}( \langle n\rangle_{\omega_{fi}}+1)=
\lambda^{ \uparrow Q} \exp( \hbar \omega_{fi} /k_B T)$. 
At temperatures $T >> \hbar \omega_{fi} /k_B$ the factor $ \langle n\rangle_{\omega_{fi}}$ 
becomes  $ \sim k_B T / \hbar \omega_{fi}$,  and the rates 
$\lambda^{ \downarrow Q}$, $\lambda^{ \uparrow Q}$  practically coincide with 
$\lambda^{noise}_{fi}$ of Eq. (17).  
\\ \indent
According to Eqs. (8) and (17),  the friction force and the 
transition rate are completely determined by the spectral density $J(\omega)$. 
In the case of Ohmic dissipation   $J(\omega) = \gamma \omega $,   
and the transition rate of  Eq. (17) becomes
\begin{equation}
\lambda_{fi}^{ \Omega}  = \frac{2  }{\hbar^2} \vert K_{fi} \vert^2
\gamma k_B T  ~~.
\end{equation}
The memory function  defined by Eq. (8)  is singular, reducing
to a delta function,   $\Gamma(t) = 2 \gamma \delta(t)$, and therefore
the calculation of the friction force requires a special discussion.
For a classical particle with the velocity $\dot{Q}$, the singularity is removed 
assuming that the cutoff frequency $\omega_c$ of $\mu_\omega$ is such that
$\dot{Q}/\omega_c$ is very small, and the delta
function is an  approximation for a smooth function $\zeta (t)$ \cite{zw}, 
peaked at $t=0$ but non-vanishing for $\vert t \vert \neq 0$.
The friction force obtained from Eq. (8) using these assumptions is
\begin{equation}
f_{\rho} (t)= - 2 \gamma \dot{Q}_\rho (t) \int_{0}^t \zeta(t-t')  dt' =  
- \gamma \dot{ Q}_\rho (t)  ~~.
\end{equation}  
If $\delta(t)$ is supposed to represent the limit of a smooth
function defined over the interval $(- \infty,0]$, then the friction force 
increases by a factor of 2 with respect to the estimate of Eq.
(23), being too strong to ensure thermalization in a classical system. 
The power dissipated by the friction force of Eq. (23) at $T=0$, when the 
noise disappears, is\footnote{For comparison, the power dissipated in the
quantized environment, by a spontaneous decay $\vert E_f \rangle \rightarrow
\vert E_i \rangle$, is $A_{fi}(E_f-E_i)=2 \gamma \vert \langle E_f \vert
[H_0,K] \vert E_i \rangle \vert^2/ \hbar^2$.}
\begin{equation}
\frac{dE_{diss}}{dt} \equiv  -  \frac{ d Tr( H_0 \rho)}{dt} =
  \gamma (\dot{ Q}_\rho  )^2 = 
  \frac{\gamma}{\hbar^2} \vert Tr( \rho [H_0,K]) \vert^2 
~~.
\end{equation}  
\indent
Though convenient in calculations, the linear dependence of $J(\omega)$ on 
$\omega$ over the whole spectrum is unrealistic, and it is useful to 
study also a physical situation of interest.  Among the various bilinear
coupling systems \cite{ford}, particularly important is the case 
of a charged particle interacting with the electromagnetic field. 
Decoherence produced by spontaneous emission of photons from cold two-level 
atoms during  tunneling through a laser-induced potential
barrier was analysed in \cite{japha}, while a quasiclassical treatment
of the bremsstrahlung spectrum for a tunneling charge is presented   
in  \cite{dyakonov}.
Let us consider here the case of a quantum charged particle interacting
with the blackbody radiation field. In the dipolar approximation, the 
interaction  Hamiltonian has the form $H_{noise}^{rad} =
- \vec{d} \cdot \vec{E}=-d_x E_x - d_yE_y-d_zE_z$ where $\vec{d}= e
\vec{r}$ is the electric dipole operator of the particle, and
$\vec{E}$ is the radiation electric
 field. For isotropic radiation  $e \vec{E}$  represents a stochastic force
with 0 mean, such that $H_{noise}^{rad}$  is a
sum of three terms, each of the form $- K \xi$ assumed above.  
If the components of the stochastic force are uncorrelated,  the
FDT is expressed by $e^2  \ll E_i(t)E_j(t')\gg=\delta_{ij} k_B T \Gamma(t-t')$. For
$t=t'$, the left-hand side of this equality  can be related to 
the total spectral energy density $u_\omega$, 
\begin{equation}
 \ll E_i^2(t)\gg = \frac{4 \pi}{3} \int_0^\infty d \omega u_\omega~~,
\end{equation}
and if $\Gamma(t)$ in the FDT has the form assumed in Eq. (8),  then
\begin{equation}
J (\omega) = \frac{2 \pi^2 e^2}{3  k_B T} \omega  u_\omega~~.
\end{equation}
For $u_\omega$ the realistic expression $ u_\omega = \hbar \omega^3
 \langle n\rangle_\omega / \pi^2 c^3$ can be considered,  which accounts for the effects of
the second quantization of the radiation field.
In this case, the transition rate provided by Eq. (17) is
\begin{equation}
\lambda_{fi}^{ rad}  = A_{fi}  \langle n\rangle_{ \omega_{fi} }
\end{equation} 
where
\begin{equation}
A_{fi} =   \frac{4 e^2}{3 \hbar c^3    }
 \vert \vec{r}_{fi} \vert^2   \omega_{fi}^3  
 \end{equation}
denotes the Einstein coefficient for spontaneous emission. Therefore, 
$\lambda_{fi}^{ rad}$ is the usual rate of the
stimulated  transitions in thermal  radiation field \cite{dirac}.
The memory function vanishes when $T=0$, but
at high temperatures  it becomes proportional
to the second derivative of the delta function,
$
\Gamma(t)  = -  4 e^2 \ddot{\delta}(t) /{( 3 c^3)}
$,
and the corresponding friction force 
\begin{equation}
\vec{f}_{rad}  = \frac{ 2 e^2}{3 c^3 } \frac{ d^3  \langle  \vec{r}\rangle}{dt^3},
\end{equation}
is the classical radiation reaction \cite{jack}.
\\[.5cm]
{\bf IV. Numerical results for a quantum two-state system } 
\\[.5cm] \indent
The model of a quantum two-state system (TSS) interacting with the environment 
appears in many physical situations \cite{legg}. Usually it is associated
with the relaxation of a $1/2$ spin system, but it was adapted  
also to study the environmental  effects on the quantum coherence oscillations 
(QCO) in one-dimensional double-well  potentials. \\ \indent
A quantum TSS can store one bit of information (qubit), and represents the 
main component in the design of a quantum computer. The
influence of the environment is crucial in this case, because an
accurate execution of the logical operations \cite{barenco} requires a  
conditional, but unitary evolution of the qubit states. 
According to  Cirac and Zoller, \cite{cz} a first generation of quantum logic
circuits could be created using cold ions confined in a linear trap.
The qubits are reprezented by two-level
systems of internal electronic states, while the conditional dynamics
required by the logical operations is implemented by their entanglement
using the collective states of the center of mass (CM) vibrations.  
Therefore, the operation time of such computing devices
is limited essentially by the occurrence of spontaneous transitions between the 
electronic states, and the environmental decoherence of the CM 
vibrational dynamics \cite{huges}. \\ \indent
The present numerical estimates concern the quantum TSS consisting of the 
ground ($\vert 0\rangle$) and the first excited state ($\vert 1\rangle$), for the CM 
vibrations of a single ion trapped in a one dimensional harmonic oscillator 
potential.
Within this restricted two-state Hilbert space,  the
unperturbed Hamiltonian and the coupling operator
can be expressed in terms of the Pauli spin 
matrices
$\sigma_x = \vert 1\rangle \langle 1 \vert - \vert 0 \rangle \langle 0 \vert$,
 $\sigma_y = \vert 1\rangle \langle 0 \vert + \vert 0 \rangle \langle 1 \vert$,
$\sigma_z = i(\vert 0\rangle \langle 1 \vert - \vert 1 \rangle \langle 0 \vert)$.
This representation is chosen such that  ${ H}_0 = \Delta \sigma_x /2$,
and $K= Q \sigma_z $, simulating a coupling linear in the CM  coordinate
or momentum. For simplicity, the spectrum of the environmental noise will be
considered flat (white noise),  having the density $J(\omega) = 
\gamma \omega$. In this case  Eq. (9)  becomes 
\begin{equation}
i \hbar \partial_t  \rho = [H_0 
 -Q {\cal F}(t) \sigma_z , \rho]~~, ~~
\end{equation}
with $ {\cal F}(t) = \xi(t) - \gamma \Delta Q P_y / \hbar$.
Denoting by  $\vec{P} \equiv \{P_x,P_y,P_z \} $ the polarization
vector, the density operator can be written as
$
{\rho} = ( { I} + \vec{P} \cdot \vec{\sigma} )/2 
$
and Eq. (30) takes the explicit form
\begin{eqnarray}
\hbar \dot{ P}_x & = & 2 Q {\cal F}(t) P_y \nonumber \\ 
\hbar \dot{ P}_y & = & - \Delta P_z -2 Q { \cal F}(t) P_x \\
\hbar \dot{ P}_z & = & \Delta P_y  \nonumber
\end{eqnarray}
 It can be easily seen that $\vec{P} \cdot \vec{P}$ is a constant of motion, 
and  along each trajectory $\vec{P}(t)$ the  purity of the states is 
preserved.  However, the evolution from pure to mixed states may appear for 
$
\rho_{av} =  \ll {\rho}\gg \equiv ( { I} +  \ll \vec{P}\gg \cdot \vec{\sigma} )/2 
$
and is described  by the dynamics of the average polarization vector
$ \ll \vec{P}\gg$. The eigenvalues $w_0$, $w_1$ of $ \ll \rho\gg$ are related to
$ \ll \vec{P}\gg$ by
\begin{equation}
w_0 = \frac{1+ \vert  \ll \vec{P}\gg \vert}{2}~~,~~
w_1 = \frac{1- \vert  \ll \vec{P}\gg \vert}{2}
\end{equation}
and determine the entropy
\begin{equation}
S = - (w_0 \ln w_0 + w_1 \ln w_1)~~. 
\end{equation}
\indent
In the numerical calculations $\Delta= \hbar \omega =19.74$ 
$10^{-9}$ eV was considered, corresponding to a trap frequency  
$\omega= 30$ MHz \cite{poyatos}.  The environmental  
decoherence of the CM wave functions in a Cirac-Zoller logical gate 
is due to  the heating of the ion vibrational motion, or to the
random phase fluctuations of the laser fields \cite{huges}. 
To simulate the global effect of these external factors an effective coupling
parameter  \cite{legg} $\alpha \equiv \gamma Q^2/ \hbar = 10^{-4}$ was chosen.
This value is somehow large, because it corresponds
to a spontaneous decay rate (Eq. (21)), $A_{fi}= 2 \alpha \omega = 6 $ 
kHz, greater than the values of practical interest \cite{huges}. 
However, it is small enough to be relevant for understanding the behavior of
the system at weak damping. Further decrease of $\alpha$ is not expected to
bring new qualitative features, though the numerical calculations may become
difficult by the increase of the computer time. The temperature was fixed
at  $T = 1$ mK, when the thermal energy $k_B T = 86.2 $ 
$10^{-9}$  eV is relatively high compared to $\Delta$, and 
the  corrections due to the quantization of the environment's degrees
of freedom may be considered small. The choice of a high temperature makes
the two-state approximation too restrictive for a complete description of the
vibrational dynamics, but despite this aspect, it remains suggestive because a
linear coupling in the CM coordinate or momentum produces transitions only
between  consecutive levels.
\\ \indent
The numerical integration of Eq. (31) was performed using the D02BAF
routine of the NAG library \cite{nag} using a time step $dt=0.658$ ns.
The initial condition $\vec{P} (t=0) =\vec{P}_0$  was chosen to represent
a pure state, and the average of the density operator was calculated over
an ensemble of $N_t = 1000$ trajectories. For each trajectory
 the noise $\xi(t)$ at the moment $t_n=ndt$  was expressed by
$
\xi(t_n) = R_n   \sqrt{  2 k_B T \gamma/dt} 
$
where $\{ R_n$, $n=1,2,3,.... \}$  is a sequence of Gaussian pseudo-random
numbers with $0$ mean and variance 1. This choice ensures the FDT in discrete
form, $ \ll  \xi (t_j ) \xi (t_k ) \gg = 2 k_B T \gamma \delta_{t_j t_k} / dt $. 
\\ \indent
The vector $\vec{P}_0$ was taken of the form  $\vec{P}_0 =(
\cos \Phi,0, \sin \Phi)$ with $\Phi =0$ and $ \pi/2$. These
values of $\Phi$ correspond to an initial preparation of the system
in the upper stationary state  $\vert 1\rangle$, and in the non-stationary 
linear superposition  $( \vert 1\rangle + i \vert 0\rangle )/ \sqrt{2} $, respectively.  

\begin{figure}
\begin{picture}(100,215)(0,0)
\put(-10,-5){\includegraphics[width=5in,height=3.1in]{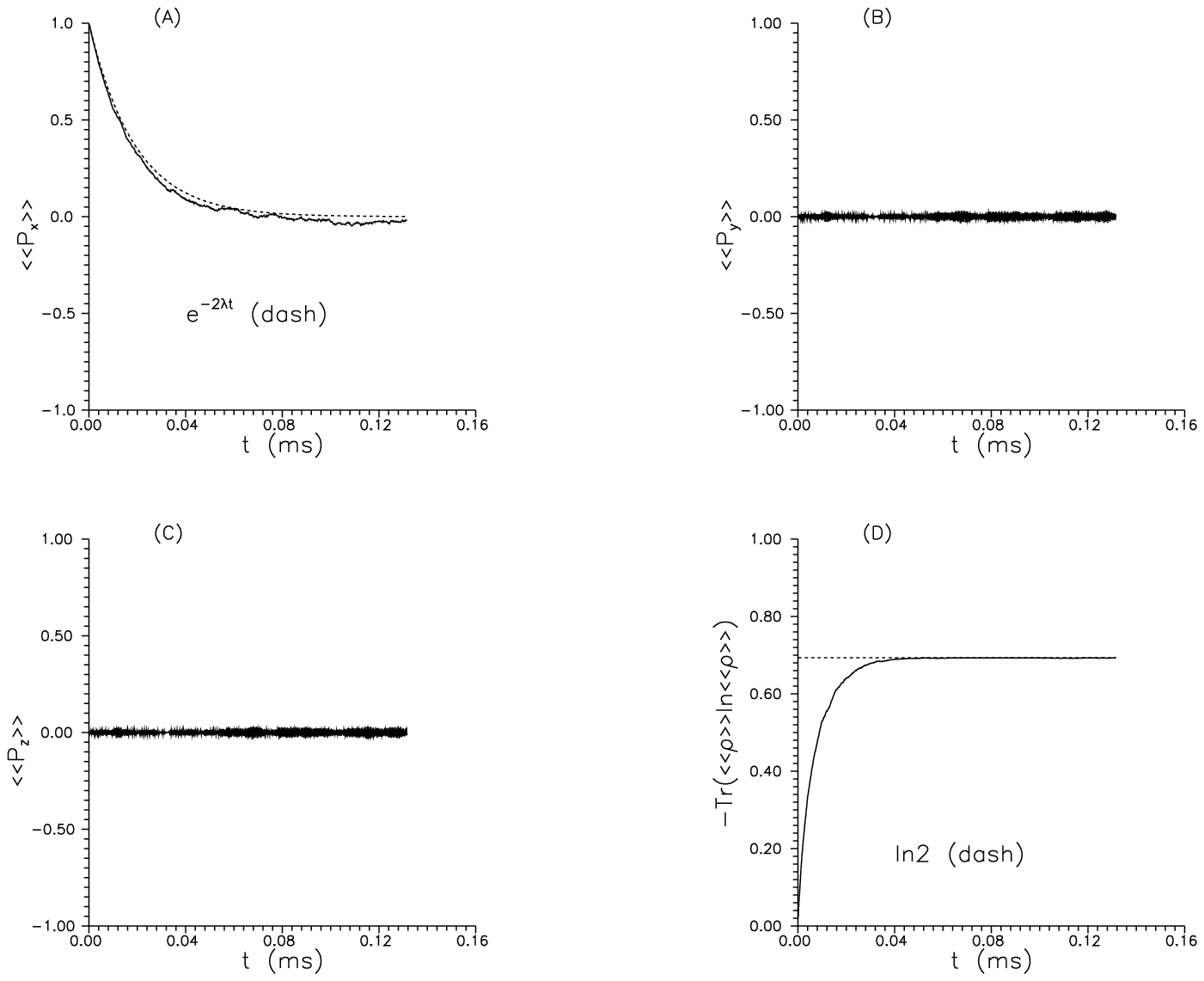}}
\end{picture}
\vskip.2cm
{\small Fig. 1. Ensemble average of the polarization vector (A)-(C),
and the entropy (D), as a function of time. The initial
condition  $\vec{P}_0= (1,0,0)$ corresponds to the upper
stationary state $\vert 1\rangle$.}
\end{figure}

 The evolution of the average polarization vector and entropy obtained
when $\Phi=0$ are presented in Fig. 1.  The average  $ \ll P_x\gg$  (Fig. 1 (A), 
solid line) is very close to the exponential $\exp( - 2 \lambda t)$ (Fig. 1 
(A),  dashed line) with $\lambda =  2 \alpha k_B T / \hbar= 26.2$ kHz
given by Eq. (22).   
This result is natural, because if the the occupation probabilities $v_0= 
 \langle 0 \vert \rho_{av} \vert 0\rangle$ and $v_1=  \langle 1 \vert \rho_{av} \vert 1\rangle$  depend
on time according to the  "phenomenological"  rate  equations
\begin{equation}
 \dot{v}_0 = \lambda^{\downarrow}v_1 - \lambda^{\uparrow} v_0 ~~,~~ 
\dot{v}_1 = - \lambda^{\downarrow} v_1 + \lambda^{\uparrow} v_0
\end{equation}
then when $\lambda^{\uparrow} \approx \lambda^{\downarrow}
\approx \lambda$, the relaxation rate of $ \ll P_x\gg= v_1-v_0$ is $2 \lambda$. 
The same rate characterizes the decrease of the energy 
$E= Tr( \rho_{av} H_0 )= \Delta  \ll P_x\gg/2$. 
\\ \indent
The averages $ \ll P_y\gg$,  $ \ll P_z\gg$ presented in Fig. 1 (B),(C), fluctuate 
around zero,
and the main contribution to the variation of the entropy  (Fig. 1 (D), solid 
line) is due to $ \ll P_x\gg$. The final value of $S$ is very close to
$\ln 2$,  (Fig. 1(D), dashed line), the
result expected when there is complete decoherence and $w_0=w_1=1/2$. 
Therefore, the pure state $\vert 1\rangle$ evolves to an incoherent 
mixture of the two states, $\vert 0\rangle$ and $\vert 1\rangle$. 
It is interesting to note that the dissipative friction force
has a relatively small contribution to the energy decrease.
This aspect becomes more clear when $\Phi= \pi/2$.  In this case, without
environment coupling ($\gamma=0$), $P_x$ is a constant, while
$P_{y,z}$ perform "QCO", $P_y (t) = \sin \omega t$, $P_z (t) = \cos
\omega t$. With coupling, the evolution of the average polarization vector 
and of the entropy are  presented in Fig. 2. The average $ \ll P_x\gg$ (Fig. 2 (A)) has small fluctuations near the initial
value, while  the oscillation amplitudes of  $ \ll P_y\gg$ and $ \ll P_z\gg$
decrease exponentially in time\footnote{due to the noise, because for
a non-stationary initial state  Eq. (14) yields a transport equation,
in our case with the solution $ \ll P_z\gg+i \ll P_y\gg=e^{- \lambda t}
e^{i \Omega t}$, $\Omega=\sqrt{\omega^2- \lambda^2}$.}
(Fig. 2 (B), (C), dashed line) and asymptotically the coherence is lost
(Fig. 2 (D)).

\begin{figure}
\begin{picture}(100,215)(0,0)
\put(-10,-5){\includegraphics[width=5in,height=3.1in]{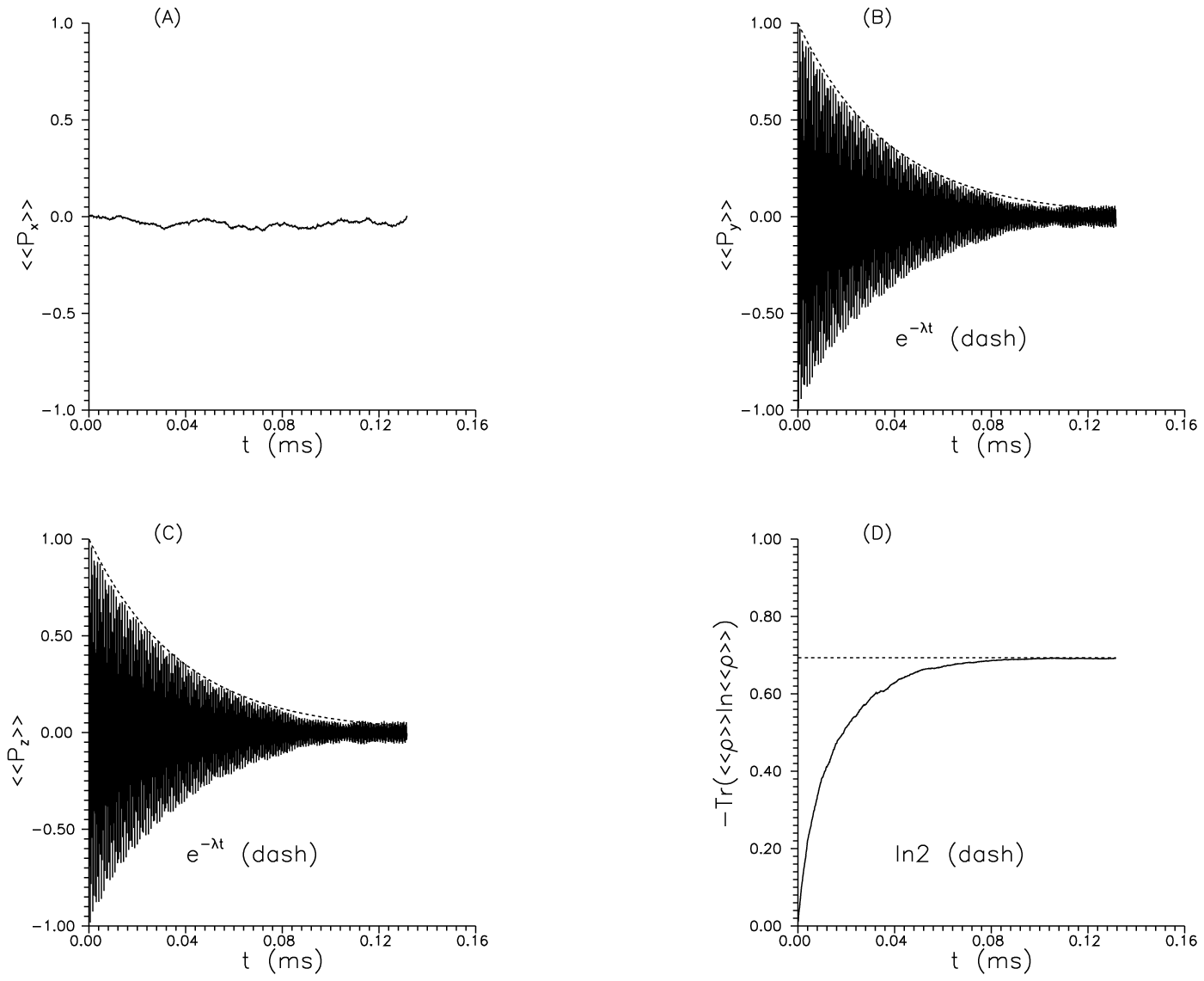}}
\end{picture}
\vskip.2cm
{\small Fig. 2.  Ensemble average of the polarization vector (A)-(C),
and the entropy (D), as a function of time. The initial condition
$\vec{P}_0= (0,0,1)$ corresponds to the non-stationary superposition
$( \vert 1\rangle + i \vert 0\rangle )/ \protect \sqrt{2} $.}
\end{figure}

In the calculations presented above the average excitation energy 
$E_x=\Delta/2+ Tr( \rho_{av} H_0)$ changes in time due to the
combined effect of noise and friction. When $T >0$ a well-defined
separation between these two contributions is not possible due
to the non-linearity, but a measure of the energy which is dissipated by
the friction force  alone is the ensemble average of $E_{diss}$ from Eq. (24).
For the TSS considered above this average has the form
$ \ll E_{diss}\gg = 0.5 \Delta  \ll P_x^d\gg $, where $ \ll P_x^d\gg$ is the
ensemble average of the component $P_x^d$ obtained  by  integrating the
equation
\begin{equation}
\frac{ d P_x^d }{dt} = \kappa(t)~~,~~\kappa=\frac{2
\gamma}{\Delta} 
( \dot{Q}_\rho )^2= A_{fi}  P_y^2~~.
\end{equation}

\begin{figure}
\begin{picture}(100,175)(0,0)
\put(-10,-5){\includegraphics[width=5in,height=2.5in]{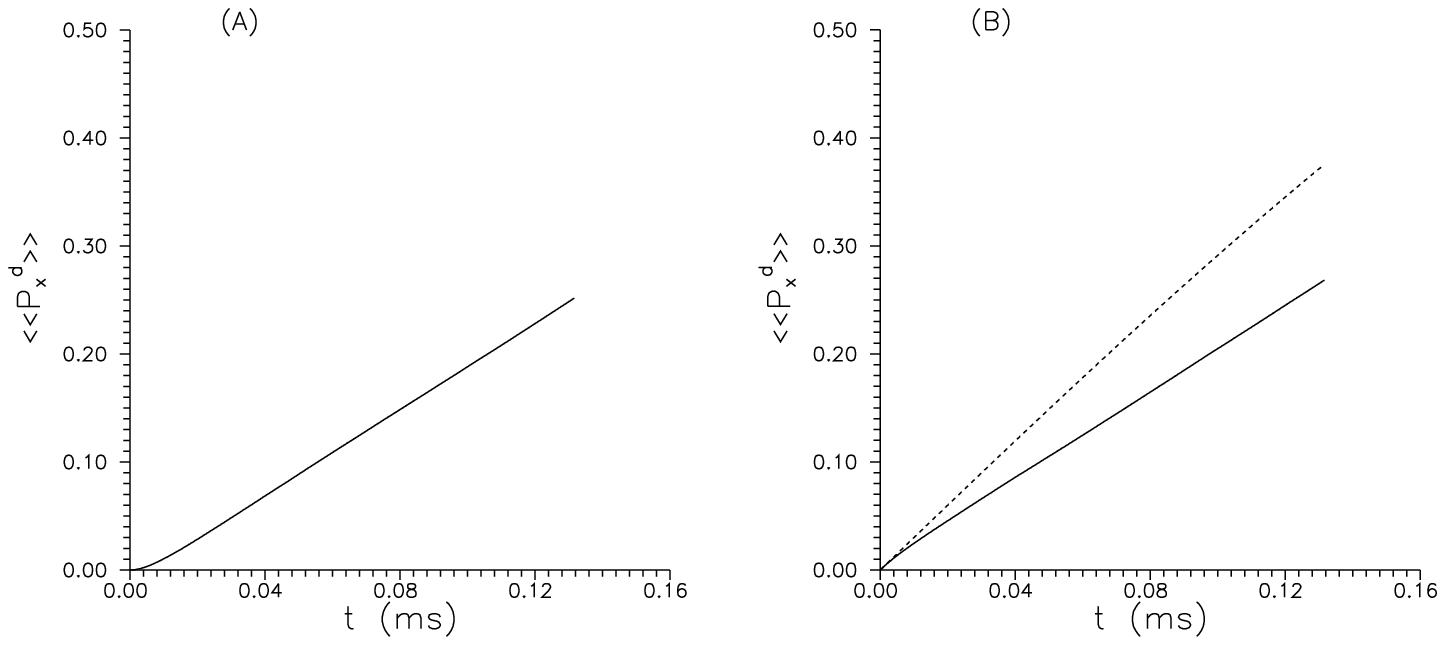}}
\end{picture}
\vskip.2cm
{\small Fig. 3. The ensemble average $ \ll P_x^d\gg$ as a function of time  when 
$\vec{P}_0= (1,0,0)$ (A)  and $\vec{P}_0= (0,0,1)$ (B, solid line). 
The short-time behavior of $P_x^d$  calculated without noise (B,
dashed line).}
\end{figure}

The  averages $ \ll P_x^d\gg$ corresponding to the trajectories presented in 
Fig. 1 and 2 are pictured in  Fig. 3 (A) and (B), respectively,
by solid line. After a short transitory time, $ \ll  \kappa \gg$ becomes 
practically a  constant  $\kappa_e \sim 2$ kHz, independently of the initial 
condition.  This constant  may provide a measure of the 
effective coupling between the quantum system and the thermal environment,
because at equilibrium  the average energy  dissipated by friction 
should be equal to the average energy which is transferred to the system by 
the thermal  noise.  \\ \indent
The short time behavior of $ \ll P_x^d\gg$ is determined  by the frictional forces
and the initial state. If $\Phi = \pi/2$ and Eqs. (31) are integrated
considering $\xi(t)=0$, then for small times $P_x^d$ increases linearly in
time   (Fig. 3 (B), dashed  line), with a slope $\kappa_0 = A_{fi}/2=3$ kHz. 
Asymptotically this linear dependence changes to exponential.
The numerical calculations show that for  $ t > t_s 
\sim 0.4 $ ms,  $P_x^d(t) \approx {\cal P}_s^1(t)$, ${\cal P}_s^n(t) =
n-[n-P_x^d(t_s)] \exp[-A_{fi} (t-t_s)]$, 
and the QCO oscillation amplitude is 
$a_{QCO}(t) \sim  a_{QCO}(t_s) \exp[-A_{fi}(t-t_s)/2]$. \\ \indent
When $\Phi=0$ the friction force vanishes, and without noise
($\xi(t)=0$),  the system  remains in  the upper stationary  
state,  without dissipation. However, if $\Phi$ is very small, but not 0, 
then $a_{QCO}$ increases, attains the maximum, and then decreases,   
such that after a transitory time $t_s$, $P_x^d$ has the
exponential behavior noticed above,  $P_x^d (t) \approx {\cal P}_s^2 (t)$. 
\\[.5cm]
{\bf V. Summary and Conclusions } 
\\[.5cm] \indent
A quantum system in interaction with the surrounding environment
has a complex dynamics, which cannot be described only by linear unitary
transformations in the Hilbert space between pure states. 
When the  environment is classical, the dynamical equations  can be obtained
using a time-dependent variational principle.  In this work the environment's
degrees of freedom have been simulated by classical harmonic oscillators,
while the dynamical variables of the  quantum system are two non-hermitian 
"square root operators" $\eta$,  $\eta^\dagger$, defined by a Gauss-like
decomposition of the density matrix. \\ \indent
If the coupling is bilinear, the  evolution of the density operator is 
described by  a Liouville equation (Eq. (6)) with an effective Hamiltonian
containing two additional terms. One of these terms ($\sim \xi$) is due to the
external forces,  while the other ($\sim f_\rho$) to the retarded backreaction
of the  environment on the quantum system, and is non-linear.  Therefore, the
backreaction term preserves the purity of the states, but may restrict the
validity of the superposition principle. \\ \indent
When the environmental oscillators are distributed within a statistical
ensemble, the two additional force  terms correspond to the noise
and friction, being related  by the fluctuation-dissipation 
theorem.  In this case, the density operator has a Brownian trajectory
corresponding to the evolution from pure to mixed states,
described by the ensemble average.
\\ \indent 
The ensemble average involved in the evaluation of the 
transition rate between eigenstates can be obtained analytically (Eq. (17)) 
if the noise is treated as a perturbation and the friction is neglected. 
For comparison, this average was calculated also when the environment
is described by the quantum, rather than classical statistical mechanics, 
using the partial tracing in the whole Hilbert space (Eq. (21)). As expected, 
the two averages become close when the thermal energy is greater than
the transition energy.  \\ \indent
The perturbative transition rates and the frictional forces are given 
explicitly for an Ohmic environment
and for the physical situation of the blackbody radiation surrounding  a 
quantum charged particle. This latter example shows that, by a suitable
choice of the environmental spectral density, it is possible
to recover the stimulated transitions rate provided by the Fermi's golden
rule. 
   \\ \indent
A non-perturbative solution of the stochastic non-linear Liouville equation 
was obtained numerically  for the case of a two-state system coupled to an
Ohmic environment. The two states correspond to the ground and first 
excited state of the CM vibrations for an ion confined in a harmonic trap,
at a temperature of 1 mK. Two different initial conditions have been 
considered, one corresponding to the upper eigenstate of $H_0$, ($\Phi=0$), 
and the other to an  eigenstate of the "coordinate" operator $K$,
 ($\Phi=\pi/2$). The upper energy eigenstate 
is also a linear superposition between the two 
"localized" eigenstates of 
 $K$, separated by $\Delta_K = 2Q$.  
For this state,  complete decoherence appears  
when the excitation energy becomes half of the initial value (Fig. 1), 
after a time 
$ \sim \tau_D= (2 \lambda)^{-1}= \hbar^2 /( \gamma k_B T \Delta_K^2) $,
in agreement with the previous estimates \cite{zurek}. 
 The decoherence time of the "localized" non-stationary superposition 
 is greater by a factor of 2, but the excitation energy 
in this case fluctuates near its initial value (Fig. 2).  \\ \indent
 At high temperatures
the power dissipated by friction is relatively small (Fig. 3), with an  
equilibrium value which is practically independent of the initial state. 
Without noise, the classical environment produces dissipation only if 
the  initial state is non-stationary  (e.g. $\Phi>0$). For 
asymptotic times this frictional dissipation resembles closely the 
spontaneous decay obtained when the environment is quantized.

\end{document}